\documentclass{emulateapj}

\shorttitle{The Stellar Populations of NGC 4522}
\shortauthors{Crowl \& Kenney}

\begin{document}

\title{The Stellar Population of Stripped Cluster Spiral NGC 4522: A
  Local Analog to K+A Galaxies?}
\author{Hugh H. Crowl}
\author{Jeffrey D.P. Kenney}
\affil{Department of Astronomy, Yale University, New Haven, CT 06520}
\email{hugh@astro.yale.edu}

\begin{abstract}
  
  We present observations of the stripped Virgo Cluster spiral
  NGC~4522, a clear, nearby example of a galaxy currently undergoing
  ISM-ICM stripping. Utilizing SparsePak integral field spectroscopy
  on the WIYN 3.5m telescope and GALEX UV photometry, we present an
  analysis of the outer disk ($r > 3$ kpc) stellar population of this
  galaxy, beyond the HI and H$\alpha$ truncation radius. We find that
  the star formation in the gas-stripped outer disk ceased very
  recently, $\sim 100$ Myr ago, in agreement with previous claims that
  this galaxy is currently being stripped. At the time of this
  stripping, data and models suggest that the galaxy experienced a
  modest starburst. The stripping is occurring in a region of the
  cluster well outside the cluster core, likely because this galaxy is
  experiencing extreme conditions from a dynamic ICM due to an ongoing
  sub-cluster merger. The outer disk has a spectrum of a K+A galaxy,
  traditionally observed in high-redshift cluster galaxies.  In the
  case of NGC~4522, a K+A spectrum is formed by simple stripping of
  the interstellar gas by the hot intracluster medium.  These data
  show K+A spectra can be created by cluster processes and that these
  processes likely extend beyond the cluster core.
\end{abstract}

\section{Introduction}

K+A galaxies \citep{dg83}, originally discovered in clusters at high
redshift, differ from typical elliptical, spiral or irregular galaxies
in an important way. These galaxies have spectra characterized by
strong Balmer absorption lines and no significant emission from
ongoing star formation. While the formation mechanisms of K+A galaxies
are still unknown \citep{tran03,poggianti04a,bekki05,goto05}, these
spectra are generally interpreted as the signature of recent cessation
of star formation (however, c.f.  \citealp{burstein05}).
\citet{shioya02} show through star formation models that such
cessation and passive evolution of the already-existing population can
form a K+A spectrum.  While the interpretation of these spectra as the
result of temporally truncated star formation is largely accepted, the
cause of the termination of star formation is not settled.

Such catastrophic interruption of star formation seems likely to be caused
by external processes. Events such as mergers, galaxy harassment, and
ram pressure stripping all have the potential to terminate star
formation. While K+A galaxies have been chiefly observed in clusters,
observations of K+A spectra in lower density environments (i.e.
\citealp{zabludoff96,tran04,goto05}) show that they are not unique to
the cluster environment. What role does environment play in the
creation of these galaxies?

In the Virgo cluster, many spirals have truncated H$\alpha$ disks
\citep{koopmann04}, suggesting that ram pressure stripping has cut off
the star formation in the outer disk. These galaxies have normal star
formation inside a truncation radius, with no detected star formation
at larger radii. The truncation location varies from galaxy to galaxy,
with some galaxies moderately truncated inside $0.8 R_{25}$ and others
more severely truncated inside $0.4 R_{25}$. In these galaxies, it is
possible to measure age-sensitive absorption lines from the stellar
populations beyond the truncation radius without contamination of
emission lines from star-forming regions.  The ages of the youngest
stellar population in the outer disk tells us how much time has
elapsed since the last epoch of star formation and, therefore, the
time elapsed since the gas was removed. This, combined with the
galaxy's location in the cluster, can tell us where in the cluster
galaxies are stripped of their star-forming gas.

NGC~4522 is one of the clearest examples of ongoing ISM-ICM stripping
observed in the nearby Virgo cluster. It is a highly inclined,
0.5L$_*$ Sc galaxy $3\fdg3$ ($1.3 r_{200}$) south of M87, in a region
of the cluster with modest ICM density (as measured from X-ray
emission from \citealp{bohringer94}).  There are several lines of
evidence that NGC~4522 is experiencing ongoing ICM pressure. Despite
an undisturbed stellar disk, the HI and H$\alpha$ emission are
spatially truncated in the disk at $r$=3 kpc=0.4 $R_{25}$ and
significant amounts of extraplanar HI and H$\alpha$ emission exist on
one side of the disk close to the truncation radius
\citep{kk99,kenney04}.  A ridge of highly polarized radio continuum
emission is observed in the disk opposite the extraplanar gas,
suggesting compression of the magnetic field lines at the leading edge
of the ISM-ICM interaction \citep{vollmer04}. These data are
consistent with a galaxy currently undergoing ram pressure; comparison
of simulations and observations \citep{vollmer06} suggest that we are
observing this galaxy at a time close to peak pressure.

 \begin{figure*}
\includegraphics[width=7in]{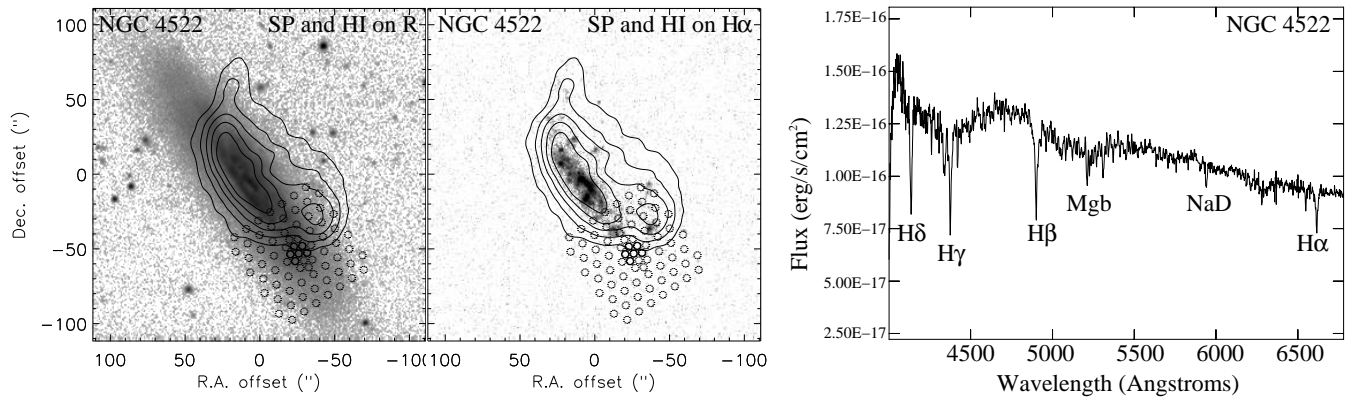}
 \caption{Left: $R$ band image of NGC~4522, with the position of the
   SparsePak fibers superimposed over the image. Middle: H$\alpha$
   image of NGC~4522, with HI contours \citep{kenney04} and the
   SparsePak footprint overlayed. In both figures, the outlines of the
   fibers used to form the composite spectrum are bold. Right: Optical
   spectrum, resulting from the combination of several fibers just
   beyond the gas truncation radius.}
 \label{images}
 \end{figure*}

\newpage

 \section{Observations}

\subsection{Optical Spectroscopy}

NGC~4522 was observed with the SparsePak integrated field spectrograph
\citep{bershady04} on the WIYN 3.5m telescope during the night of
March 29-30, 2003. SparsePak has 75 object fibers, each with a
diameter of $4\farcs7$, spread over an $80\arcsec$ x $80\arcsec$
region of the sky. The fiber spacing for most of the array is
$10\farcs3$ , except for a denser central core where the fibers have a
center-to-center spacing of $5\farcs6$ The SparsePak bundle was
positioned with the majority of the fibers located beyond the
H$\alpha$ truncation radius (Figure \ref{images}). We followed
standard IRAF image reduction techniques and used the {\tt dohydra}
task to extract the spectra, determine the wavelength solution, and
subtract sky light. The data were then flux calibrated using
spectrophotometric standards. The resulting data have a resolution of
5.5 \AA (FWHM).  Following the complete reduction of the spectral
data, the spectrum from each fiber was examined to determine the
quality of the data. The H$\alpha$ truncation radius determined by
spectroscopy matches the imaging data. We combine six fibers with no
H$\alpha$ emission and high signal-to-noise just beyond the gas
truncation radius to create a higher signal-to-noise composite
spectrum (Figure \ref{images}). These spectra sample a region along
the major axis to the SW between 45\arcsec (3.5 kpc\footnote{We assume
  a distance to Virgo of 16 Mpc.})  and 60\arcsec (4.5 kpc), within
5\arcsec (380 pc) of the major axis.  These fibers all show H$\alpha$
absorption from the underlying stellar population and no apparent
H$\alpha$ emission. All stellar population results presented here are
based on this averaged spectrum.  From these data, we extract
absorption line indices based on the Lick/IDS
definitions\footnote{Note that the data are \emph{not} smoothed to the
  Lick resolution, but that we simply use the Lick feature wavelength
  definitions.} \citep{faber85}.

\subsection{GALEX Photometry}

Observations of NGC~4522 were conducted with GALEX as part of that
instrument's All Sky Survey \citep{martin05}.  NGC~4522 was observed
for $0.1\textrm{ksec}$ in both the FUV and NUV channels.  Using a
series of circular apertures chosen to match SparsePak fiber
positions, we have extracted FUV and NUV fluxes for the same 1~kpc
region along the major axis where we extract optical spectra.  We
combine these data with F606-band ($\sim V$) HST optical imaging
\citep{kenney06} to determine FUV-to-optical and FUV-to-NUV flux
ratios. Note that, because this region has been stripped of its gas
and dust, the UV fluxes and colors are minimally affected by internal
dust absorption.

\section{Results}

\subsection{Properties of the Stripped Region}

The optical spectrum of NGC 4522 is very blue, with the continuum
continuing to rise all the way to the blue spectral limit, unlike the
red spectrum of a typical elliptical galaxy. This implies that the
light from the galaxy is dominated by young stars. This observation is
in agreement with the strong Balmer absorption in the outer disk of
NGC~4522. We see strong absorption not only in H$\beta$, but also in
the higher order lines: H$\gamma$ and H$\delta$, lines characteristic
of a young, A-star-dominated population. If we compare this spectrum
with those observed at higher redshift, we find a striking similarity
to K+A galaxies.  \citet{dressler99} quantify K+A galaxies as those
galaxies with no significant star-forming
emission\footnote{\citet{dressler99} and many others use [OII]3727 as
  their indicator of star formation activity.  As our spectral
  coverage does not extend blueward of 4000 \AA, we use H$\alpha$ as
  our star formation diagnostic.} and strong H$\delta$ absorption
(EW(H$\delta$)$ > 3$ \AA). The outer disk of NGC~4522 clearly fits
this description, with no significant H$\alpha$ emission and
EW(H$\delta$)=5.2 \AA.

The UV colors of this region show a similar signature: a nearly flat
spectrum extends from the FUV band ($\lambda_{\textrm{eff}} \sim 1500$
\AA) to the F606 band ($\lambda_{\textrm{eff}} \sim 5900$ \AA).  In
particular, NGC~4522 is brighter in the GALEX FUV bandpass than in the
NUV passband by a modest but significant amount
($F_{\textrm{\scriptsize FUV}} = 3.9\times 10^{-16}$
erg/s/cm$^2$/\AA~vs.~$F_{\textrm{\scriptsize NUV}} = 3.0\times
10^{-16}$ erg/s/cm$^2$/\AA).  Because the FUV flux fades quickly in an
evolving stellar population (e.g. \citealp{bc03}), the bright FUV
emission from the outer disk implies a very young stellar population.

\subsection{Stellar Population Models}

 \begin{figure}[t]
\includegraphics[width=3.4in]{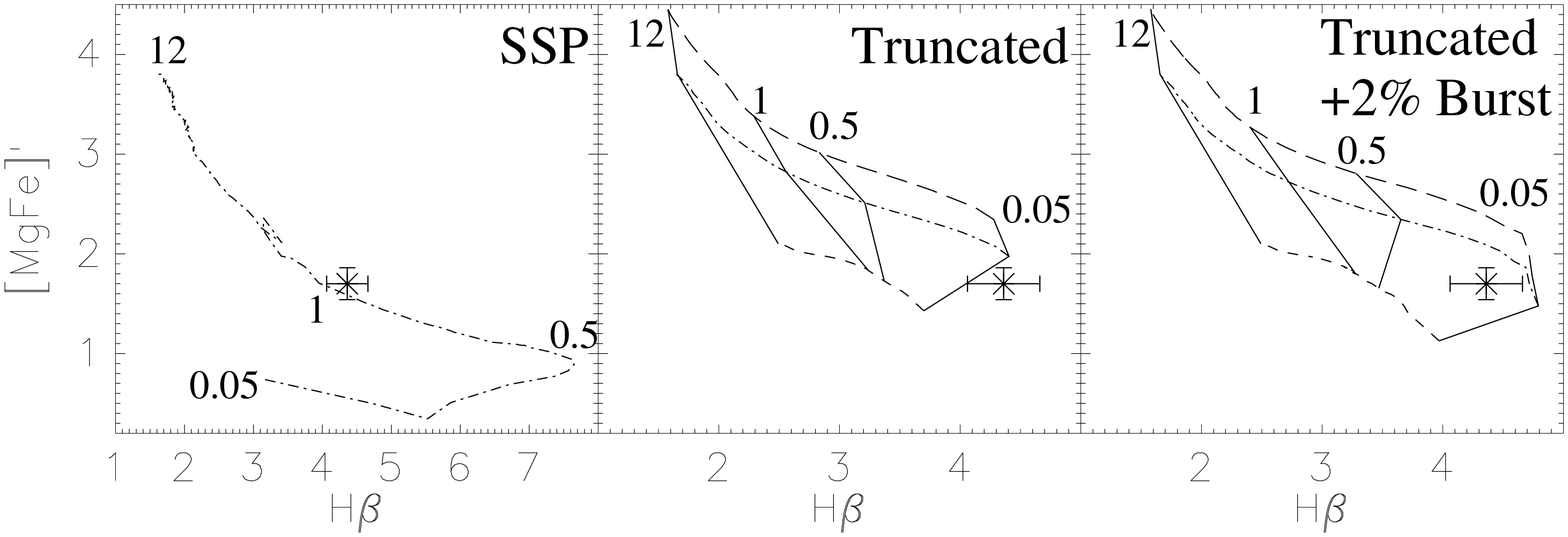}
    \caption{Index diagrams for Starburst99 models for an SSP (left),
      temporally truncated star formation (middle), and temporally
      truncated star formation with a 2\% starburst at the time of
      truncation (right). Shown are the data (star) and models (lines)
      for the age-sensitive H$\beta$ line plotted against the
      metallicity-sensitive [MgFe]$^\prime$ index. The Starburst99
      model plotted in the SSP figure (left) is for solar metallicity,
      with ages in Gyr marked along the line. In each of the other two
      figures (middle and right), the solid lines correspond to lines
      of constant truncation age, while the broken lines correspond to
      lines of constant metallicity (Z=0.008 (short dash), Z=0.02
      (dot-dash), Z=0.04 (long dash)).}
  \label{index-diag}
  \end{figure}

Spiral galaxies are typically characterized by ongoing star formation.
In the case of the stripped spiral galaxies in Virgo, it's natural to
assume a temporally truncated star formation history: roughly constant
star formation, followed by abrupt cessation of star formation and
passive evolution of the already-existing population. If one assumes a
(temporally) truncated star formation history, the youngest (and
brightest) stellar population will dominate the light, but there will
be some ``pollution'' from the older stellar population. While the
Balmer lines from the integrated light will be strong in this
population, their strength will be somewhat diluted by the older
population. In contrast, models assuming a ``single burst'' star
formation history (a so-called ``Simple Stellar Population'' or
``SSP''), will have stronger Balmer lines for $\sim 1$ Gyr, the
approximate lifetime of the Balmer-line-dominating A~stars.  This
means that, for a given set of Balmer line indices, the last epoch of
star formation in the temporally truncated model will always be
\emph{younger}\footnote{Any ``age'' referred to with respect to the
  temporally truncated star formation model is the age of the youngest
  generation of stars; \emph{not} the luminosity weighted mean age of
  the population.} than the age of the SSP model. The GALEX UV data are,
therefore, critical for constraining the age of the stellar
population. The detection of bright, FUV flux in the outer disk of
NGC~4522 demonstrates that the stellar population is very young.

Due to the presence of very young stars in the outer disk of NGC~4522
and the lack of young, metal-poor stars in empirical libraries, we
have chosen to use Starburst99 models \citep{leitherer99}, with
theoretical stellar atmospheres \citep{martins05}.  Starburst99
generates SSP models with a range of ages and the sum of several
theoretical spectra are used to model a truncated star formation
history. The spectra are then smoothed to the resolution of our data
(5.5\AA FWHM) and spectral line indices and broadband photometric
colors are extracted.  In Figure \ref{index-diag}, we plot the
strength of the age-sensitive Balmer features against the strength of
the metallicity-sensitive [MgFe]$^\prime$ index for a SSP, a set of
temporally truncated star formation models, and a set of temporally
truncated models in which 2\% of the stellar mass formed in a burst at
the end of the star forming episode. In Figure \ref{fig-galexcolors},
we plot broadband UV and optical flux ratios for same three star
formation histories.

\subsection{Time Since Outer Disk Truncation}

 \begin{figure}[t]
   \plotone{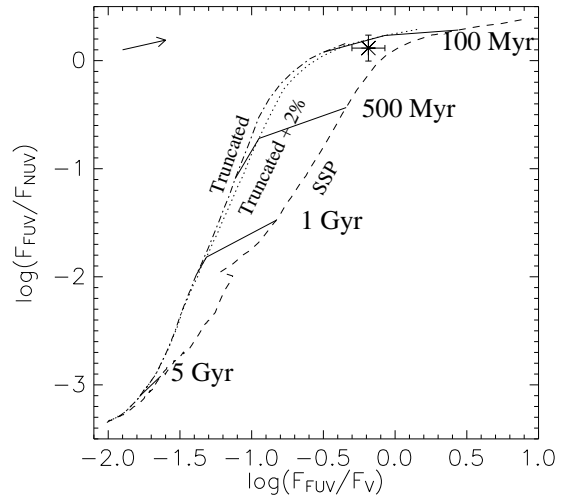}
   \caption{UV-optical colors derived for the models shown in Figure 2:
     SSP (dashed line), temporally truncated (dash-dot line), and
     truncated with 2\% burst (dotted line). The solid lines correspond
     to ages of (from top to bottom) 100 Myr, 500 Myr, 1 Gyr and 5
     Gyr. The value of the flux ratios for NGC~4522 is plotted as a
     asterisk. A representative reddening vector (determined from the
     results of \citealp{calzetti97} and using a value of E(B-V)=0.02
     from \citealp{schlegel98}) is shown in the upper left section of
     the plot.}
   \label{fig-galexcolors}
   \end{figure}

The GALEX UV photometry (Figure \ref{fig-galexcolors}) and optical
spectroscopy (Figure \ref{index-diag}) are best fit by an temporally
truncated Starburst99 model, with a cessation age of $\sim 100$ Myr.
This is based on the combination of strong Balmer absorption and
bright FUV emission in the outer disk.  Either the optical
spectroscopy or the GALEX photometry, when taken independently, can be
fit by various models. The Balmer absorption lines indicate that
either the stellar population is younger than a 50 Myr truncated star
formation model \emph{or} as old as a 1 Gyr SSP.  Truncation is clearly
the more logical star formation history, and we find the GALEX colors
are far too blue for a 1 Gyr SSP; in fact, the GALEX data is most
consistent with being truncated $\sim 100$ Myr ago, an estimate that
is relatively unaffected by reddening due to the fact that the
reddening vector is nearly parallel to the line of constant age
(Figure \ref{fig-galexcolors}). While the Balmer lines are stronger
than any simple truncated stellar population model, a modest 2\% (by
mass) starburst is enough to bring the absorption line data and GALEX
data into agreement (Figure \ref{index-diag}). Taken together, the
data exclude a SSP star formation history and show roughly constant
star formation in the outer part of NGC~4522 until $\sim 100$ Myr ago,
when there was a modest starburst, followed by the current period of
no star formation and passive stellar evolution.  This scenario is
consistent with the observations that this is a spiral galaxy
currently being stripped of its neutral gas \citep{kenney04,vollmer06}
and with H$\alpha$ observations suggesting that the inner disk has a
moderately enhanced star formation rate \citep{kenney04}.

\section{Discussion}

Our results clearly show that the cessation of star formation was
recent and, therefore, that the outer disk of NGC~4522 was recently
stripped of its star-forming gas. We would like to understand how
rapidly star formation stops during an interaction with the ICM and
how long dense molecular star-forming clouds might survive after the
low-density atomic gas is stripped. Simulations \citep{vollmer06}
compared with the HI observations \citep{kenney04} suggest that we are
observing the galaxy $\sim 50$ Myrs after peak pressure.  This
timescale is consistent with that determined from the stellar
population, suggesting that neutral gas stripping and termination of
star formation nearly coincide.

\subsection{Stripping Outside the Cluster Core}

From a simple timescale argument, the stellar populations of NGC~4522
tell us that this galaxy must be stripped locally, 0.9~Mpc ($1.3
r_{200}$) from M87, as opposed to closer to the cluster core. If we
assume that the galaxy's motion in the plane of the sky is a factor of
2 higher than the average radial velocity of a Virgo Cluster galaxy
(i.e.  we assume $v_{\textrm{\scriptsize t}} = 1400$ km/s), it would
still take the galaxy $\sim 500$ Myr to travel the $\sim 0.75$ Mpc
that separates its current location from the region where the ICM
density is high enough to strip the outer disk.  \emph{The observation
  that star formation ceased $\sim 100$ Myr ago therefore rules out
  stripping in the core and argues that stripping must be occurring
  locally.} This result is in agreement with simulations of the HI
morphology and kinematics \citep{vollmer06} which also argue for local
stripping.  Stripping so far out in the cluster shows either that gas
stripping of galaxies in clusters is easier than the simple estimates
of ram pressure imply \citep{kenney04}, or that some galaxies
experience extreme conditions due to a dynamic ICM
\citep{dupke06,heinz06}. In particular, NGC~4522 appears to be located
between M87 and the M49 subcluster, which is likely in the process of
merging with Virgo \citep{schindler99,shibata01}.  Moreover,
\citet{shibata01} observe a shock front in the ICM near the location
of NGC~4522, suggesting that the local ICM has a high velocity. This
makes it plausible that gas stripping is important for transforming
galaxies well outside the cluster core, including at the interface
between merging subclusters.

\subsection{Similarity to Higher-z K+A Galaxies}

The outer disk of NGC~4522 has strong Balmer lines and no emission,
similar to K+A galaxies observed at higher redshift. While the outer
disk of NGC~4522 shows a K+A spectrum, a global spectrum of this
galaxy does \emph{not} show the same signature, since it contains
emission lines from the star-forming inner disk and absorption lines
mostly from the stripped outer stellar disk \citep{gavazzi04}. This
makes it different from the K+A galaxies observed at higher redshift
and more similar to the ``e(a)'' class \citep{dressler99}. Spectral
modeling \citep{shioya02} has suggested that the e(a) class may evolve
into K+A galaxies, but some observations (i.e. \citealp{balogh05})
indicate that e(a) galaxies and K+A galaxies are morphologically
distinct.  There may be more than one way to make a K+A and e(a)
galaxies, but the present observations show that ram pressure
stripping \emph{can} create a K+A or e(a) spectrum, either from total
stripping (K+A) or partial stripping (e(a)). This shows that ram
pressure is effective in rapidly terminating star formation and is
capable of creating the post star formation spectra observed at higher
redshift. It may be that, in the case of Virgo, the ICM is not dense
enough to completely strip the gas from a massive galaxy in a single
passage so that a central star-forming gas disk will remain after this
stripping event ends.  In the largest clusters, the ICM density is
higher by a factor of $\sim 10$ and galaxy velocities are higher by a
factor of $\sim 2$; this means that ram pressure is $\sim 40$ times
higher in these clusters.  In such clusters, it is likely that a
galaxy like NGC~4522 would be completely stripped of its gas.  In
fact, \citet{poggianti04} have observed a striking correlation between
K+A galaxies and the locations of X-Ray structure in the Coma Cluster,
suggesting that we are observing such transformations there.  Both
these observations and our observation of NGC~4522 indicate that
cluster processes are important for the creation of some K+A spectra.
While K+A/e(a) galaxies may not all be created in clusters, it appears
that at least one route for their formation is the termination of star
formation by ISM-ICM stripping.  Moreover, it appears that stripping
can drive this transformation well outside the cluster core.

\vspace{2mm}

We gratefully acknowledge the indispensable discussions and advice of
Jim Rose, which were critical at the early stages of this paper. We
also would like to thank Pieter van Dokkum, Jacqueline van Gorkom, and
Bernd Vollmer for comments that served to clarify the discussion.
Finally, we wish to thank the referee for insightful comments and
suggestions. This research is supported by NSF grant AST 00-71251.

\end{document}